# The environments of $z < 0.3$ QSOs

R.J.Smith[1], B.J.Boyle[2], S.J.Maddox[2]

[1] *Institute of Astronomy, University of Cambridge, Madingley Road, Cambridge, CB3 0HA*

[2] *Royal Greenwich Observatory, Madingley Road, Cambridge, CB3 0EZ*

**ABSTRACT**

We have carried out an investigation of the galaxy environments of low redshift ($z < 0.3$) QSOs by cross-correlating the positions on the sky of X-ray-selected QSOs/AGN identified in the *Einstein* Medium Sensitivity Survey (EMSS) with those of $B_J < 20.5$ galaxies in the APM galaxy catalogues. At $< 5$ arcmin, we find a significant ($5\sigma$) galaxy excess around $z < 0.3$ QSOs. The amplitude of the low redshift ($z < 0.3$) QSO-galaxy angular cross-correlation function is identical to that of the APM galaxy angular correlation function, implying that these (predominantly radio-quiet) QSOs inhabit environments similar to those of normal galaxies. No significant galaxy excess was found around a 'control' sample of $z > 0.3$ QSOs. Coupled with previous observations, these results imply that the environment of radio-quiet QSOs undergoes little evolution over a wide range in redshift ($0 < z < 1.5$). This is in marked contrast to the rapid increase in the richness of the environments associated with radio-loud QSOs over the same redshift range. The similarity between QSO-galaxy clustering and galaxy-galaxy clustering also suggests that QSOs are unbiased with respect to galaxies and make useful tracers of large-scale structure in the Universe.

## 1 INTRODUCTION

The study of the QSO environments provides a useful insight into the triggering and fuelling mechanism for QSOs (Yee & Green 1984 and references therein) and can, in principle, place useful constraints on the evolution of structure in the Universe (Shanks & Boyle 1994). Due to the lack of suitable low redshift ($z < 0.3$) samples, most studies of QSO environments have centred on intermediate redshift QSOs. Yee and collaborators (Yee & Green 1984, Ellingson, Yee & Green 1991) have shown that the richness of galaxy clusters associated with radio-loud QSOs increases rapidly over the redshift interval $0.3 < z < 0.6$, with radio-loud QSOs at $z \sim 0.6$ being associated, on average, with Abell richness class 1 clusters. In contrast, the galaxy environments of radio-quiet QSOs show no evidence for any evolution in richness over these redshifts and are consistent with a much weaker field galaxy environment in this redshift range.

The difference between the environments of radio-loud and radio-quiet QSOs is borne out by studies of QSO environments at even higher redshifts. Deep CCD imaging of the fields surrounding $z \sim 1$ radio-loud QSOs (Tyson 1986, Hintzen, Romanshin & Valdes 1991) has revealed a significant over-density of galaxies, implying a rich cluster environment for radio-loud QSOs at these redshifts. However, similar studies of radio-quiet QSOs at $z \sim 1$ (Boyle & Couch 1993) reveal no galaxy excess, consistent with a field galaxy environment (albeit with large errors) for these QSOs.

Unfortunately, conclusions drawn from deep CCD imaging studies of QSO environments at intermediate redshifts are highly dependent on the correction applied to convert *observed* galaxy excess on the sky to *physical* galaxy clustering at the redshift of the QSO. Essentially this requires knowledge of the galaxy number–redshift relation $n(z)$ at the magnitude limit of the imaging study (Longair & Seldner 1979). For all but the lowest redshift QSOs studied to date ($z \sim 0.3$, $R_{\rm lim} < 22$), this invovles an uncertain extrapolation of the galaxy luminosity function and its evolution with redshift, to obtain predictions of the galaxy $n(z)$ at typical depths $R_{\rm lim} < 24$. While self-consistent modelling of the galaxy luminosity function at intermediate redshifts ($z \sim 0.6$, see Yee & Green 1984) can yield useful constraints, extrapolation to even higher redshifts becomes an increasingly difficult task.

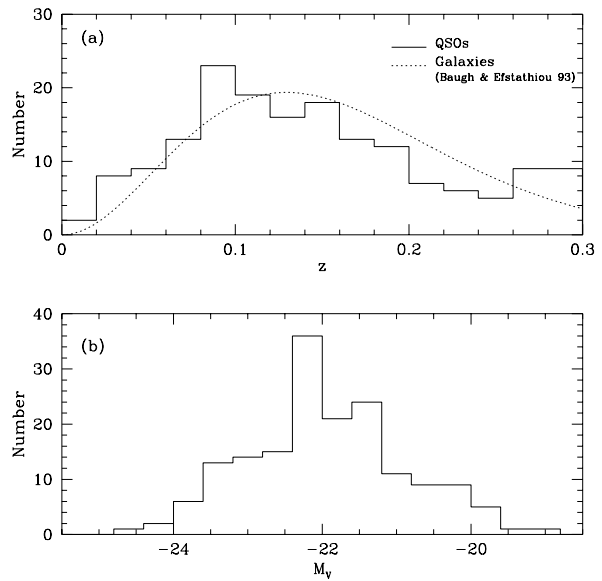

**Figure 1** (a) Comparison of the EMSS QSO and normal galaxy redshift distributions. The galaxy redshift function for galaxies down to $B_J = 20.5$ was calculated from the observationaly derived formula given by Baugh & Efstahiou (1993). Only 5 per cent of the galaxies occupy the 'tail' of the distribution, $z > 0.3$. (b) The absolute V magnitudes of the EMSS QSOs used in the present work.

In principle, more reliable estimates of the environments of QSOs could be obtained from the study of QSOs at low redshifts ($z < 0.3$), where the galaxy luminosity function is much better established. At low redshifts, the angular scales corresponding to typical cluster sizes are large; e.g. assuming $H_0 = 50\,{\rm km\,s^{-1}\,Mpc^{-1}}$ and $q_0 = 0.5$ (used throughout this paper), the characteristic Abell radius (3 Mpc) at $z = 0.15$ corresponds to 15 arcmin. At these large scales, photographic plates are more suited to the study of QSO cluster environments than CCD imaging. With the increasing availability of uniform catalogues based on machine measurements of the Northern and Southern photographic sky surveys, suitable lists of faint ($B_J < 21$) galaxies can now be generated that cover most of the sky at high $|b^{\rm II}| > 30°$ galactic latitudes.

We have therefore used such machine-based galaxy catalogues to carry out a comprehensive study of the environments of QSOs at $z < 0.3$. Our QSO list is derived from the *Einstein* Extended Medium Sensitivity Survey (EMSS), which contains a unbiased sample of over 160 X-ray-selected QSOs at $z < 0.3$. We describe both the galaxy and QSO data-sets in more detail in Section 2 and we give details of our analysis in Section 3. We discuss our results in Section 4 and present our conclusions in Section 5.

## 2 DATA

The QSOs were drawn from the 428 objects in the *Einstein* Medium Sensitivity Survey (EMSS) catalogue (Stocke et al. 1991) of X-ray selected objects, which were classified as AGN by optical identification. Despite the fact that radio-loud QSOs tend to have a slightly higher X-ray flux than the much more numerous radio-quiet QSOs (Zamorani et al. 1981), an X-ray selected sample still produces a catalogue which is representative of the general population of QSOs. Indeed analysis by Della Ceca et al. (1994) of the radio observations reported in the EMSS, confirm that only 3 per cent of the EMSS QSOs with $M_V > -24$ (i.e. all QSOs with $z < 0.3$, see below) are radio-loud. For the results quoted here, the radio-loud objects have not been removed. However, only three of the 169 objects eventually included in the sample are radio-loud, and their removal makes no measurable difference to our results. For the analysis, the EMSS QSOs were split into two sub-samples; a sample of low redshift QSOs ($z \leq 0.3$), and a 'control' sample of QSOs at higher redshifts ($z > 0.3$). After rejecting QSOs in regions of the sky where no reliable machine-based galaxy catalogue currently exists (predominantly those areas with galactic latitudes $|b^{II}| < 30°$), 169 and 156 QSOs remain in the $z \leq 0.3$ and $z > 0.3$ samples respectively. Some of the EMSS fields in which QSOs were discovered were originally targeted on Abell clusters which might introduce a slight bias into the QSO-galaxy pairs counted on these fields. However, we have determined that the exclusion of the 23 $z \leq 0.3$ QSOs affected makes no significant difference to the results presented below. This is also confirmed from the results obtained below using the 'control' sample of $z > 0.3$ QSOs.

Figure 1 presents the redshift and absolute magnitude ($M_V$) histograms for the low redshift QSO sample. Absolute magnitudes were calculated from the $V$ magnitudes listed in Stocke et al. (1991) assuming a power law spectral index, $F(\nu) \propto \nu^{-0.5}$. Over laid on the redshift histogram is a plot of the general galaxy redshift distribution function derived by Baugh & Efstathiou (1993) from redshift surveys, evaluated for a limiting magnitude of $B_J = 20.5$. The fact that there is a remarkable, and probably chance, similarity between the QSOs and galaxies (identical at the 95 per cent confidence level using the KS test), creates an important simplification to the analysis discussed in section 3.1. Note also that contamination from $z > 0.3$ galaxies is small, $\approx 5.1$ per cent.

The galaxies were magnitude limited at $B_J < 20.5$ and were drawn from two sources. The magnitude limit was chosen for consistency between the QSO and galaxy redshift distributions and reasonable reliability in the star–galaxy separation. Southern fields were taken from the APM galaxy catalogue of Maddox et al. (1990b, c) which has been sub-

**Table 1.** Values determined for the pair counts, calculated values of $w_{qg}(\theta)$ and error estimates.

| $\theta$ (arcmin) | $0.01 < z \leq 0.3$ | | | | $0.3 < z < 2.8$ | | | |
|---|---|---|---|---|---|---|---|---|
| | $N_{QG}$ | $N_{QR}$ | $w_{qg}(\theta)$ | ± | $N_{QG}$ | $N_{QR}$ | $w_{qg}(\theta)$ | ± |
| Northern Fields | | | | | | | | |
| 0.167–2 | 135 | 97.7 | 0.38 | 0.11 | 87 | 82.3 | 0.06 | 0.26 |
| 2–4 | 333 | 290.4 | 0.15 | 0.12 | 234 | 244.8 | -0.04 | 0.09 |
| 4–6 | 487 | 480.9 | 0.01 | 0.07 | 434 | 403.9 | 0.07 | 0.07 |
| 6–8 | 715 | 674.8 | 0.06 | 0.05 | 551 | 566.7 | -0.03 | 0.03 |
| 8–10 | 964 | 864.2 | 0.12 | 0.07 | 759 | 724.3 | 0.05 | 0.05 |
| 10–12 | 1082 | 1055.3 | 0.03 | 0.08 | 944 | 888.2 | 0.06 | 0.06 |
| 12–14 | 1258 | 1249.7 | 0.01 | 0.09 | 1117 | 1051.0 | 0.06 | 0.08 |
| 14–16 | 1514 | 1443.2 | 0.05 | 0.06 | 1183 | 1214.2 | -0.03 | 0.06 |
| 16–18 | 1634 | 1626.0 | 0.00 | 0.02 | 1404 | 1367.0 | 0.03 | 0.06 |
| 18–20 | 1846 | 1833.0 | 0.01 | 0.03 | 1528 | 1540.1 | -0.01 | 0.08 |
| 20–22 | 2000 | 2017.9 | -0.01 | 0.02 | 1690 | 1692.9 | 0.00 | 0.05 |
| 22–24 | 2142 | 2216.9 | -0.03 | 0.07 | 1868 | 1864.4 | 0.00 | 0.07 |
| 24–26 | 2380 | 2403.3 | -0.01 | 0.05 | 2073 | 2016.0 | 0.03 | 0.06 |
| 26–28 | 2527 | 2602.2 | -0.03 | 0.04 | 2061 | 2187.2 | -0.06 | 0.03 |
| Southern Fields | | | | | | | | |
| 0.167–2 | 76 | 55.2 | 0.38 | 0.15 | 52 | 45.0 | 0.15 | 0.35 |
| 2–4 | 184 | 167.1 | 0.10 | 0.13 | 139 | 137.0 | 0.01 | 0.21 |
| 4–6 | 293 | 276.8 | 0.06 | 0.15 | 211 | 227.1 | -0.07 | 0.06 |
| 6–8 | 384 | 390.4 | -0.02 | 0.13 | 302 | 319.8 | -0.06 | 0.08 |
| 8–10 | 517 | 503.5 | 0.03 | 0.09 | 400 | 412.9 | -0.03 | 0.15 |
| 10–12 | 601 | 613.4 | -0.02 | 0.08 | 519 | 501.3 | 0.04 | 0.12 |
| 12–14 | 761 | 724.5 | 0.05 | 0.09 | 606 | 593.1 | 0.02 | 0.05 |
| 14–16 | 873 | 843.4 | 0.04 | 0.09 | 700 | 691.5 | 0.01 | 0.06 |
| 16–18 | 954 | 950.0 | 0.00 | 0.03 | 726 | 777.3 | -0.07 | 0.11 |
| 18–20 | 1073 | 1068.4 | 0.00 | 0.07 | 832 | 873.2 | -0.05 | 0.05 |
| 20–22 | 1284 | 1169.6 | 0.10 | 0.11 | 1001 | 957.2 | 0.05 | 0.06 |
| 22–24 | 1333 | 1288.7 | 0.03 | 0.10 | 1099 | 1053.9 | 0.04 | 0.04 |
| 24–26 | 1325 | 1398.8 | -0.05 | 0.07 | 1136 | 1144.0 | -0.01 | 0.08 |
| 26–28 | 1459 | 1511.5 | -0.03 | 0.06 | 1260 | 1239.8 | 0.02 | 0.03 |

Description of column headings: $\theta$ is angular separation range to which the counts apply; $N_{QG}$ is number of QSO–galaxy pairs found; $N_{QR}$ is the number of QSO–random counts; $w_{qg}(\theta)$ is the calculated angular correlation function; ± is error on $w_{qg}(\theta)$ (see section 3.1).

ject to extensive calibration and plate matching in order to correct any plate to plate variations. The survey covers 269 UK Schmidt telescope (UKST) plates in the Southern galactic hemisphere covering the region from $21^h$ to $05^h$, and $0°$ to $-70°$. For Northern galactic hemisphere fields, the galaxy positions were taken from the APM Sky Catalogues (Irwin, Maddox & McMahon 1994), which are less heavily processed. Magnitude calibrations and star–galaxy classification have slightly higher associated errors since they are carried out on a plate by plate basis. Galaxies on 12 equatorial fields were taken from the APM Southern Sky Catalogue based on measurements of UKST equatorial $J$ survey plates. All galaxies in QSO fields with dec $> 1°$ were taken from the APM Northern Sky Catalogue, which is based on measurements of the Palomar Observatory sky survey (POSS) $O$ and $E$ survey copy plates. Two precautions were taken to ensure the reliability of the galaxy catalogue obtained from the POSS plates. First, since image quality tends to degrade towards the edge of each plate, no QSO lying more than than 2.5 degrees from any given POSS field centre was used in the analysis. Secondly, spurious galaxy identifications (e.g. stars in de-focussed regions on the original copy plate, emulsion flaws, etc.) were removed by only using objects which were classified as galaxies on both the $O$ and $E$ plate on each field. This reduces the completeness of the sample, but does not affect the result, just lowers the signif-icance. Contamination by stars classified as galaxies would change the measured amplitude, so their removal is much more important than completeness. The error in classification for the UKST plates is 5 per cent, For the POSS data, it is less well quantitfied, but is between five and ten per cent after all the precautions mentioned above have been taken. Such an error dilutes the observed amplitude linearly. i.e. reduces its value by 5–10 per cent.

## 3 ANALYSIS

### 3.1 Calculation of $w_{qg}(\theta)$

We used the angular cross-correlation function, $w_{qg}(\theta)$, to determine the strength of galaxy clustering around QSOs. For a single QSO, $w_{qg}(\theta)$ is given by:

$$w_{qg}(\theta) = \frac{N_{QG}}{N_{QR}} - 1,$$

where $N_{QG}$ is the number of galaxies found at a given separation $(\theta - \Delta\theta/2,\ \theta + \Delta\theta/2]$ from the QSO and $N_{QR}$ is the number of random points expected in the same annulus. The cross-correlation function simply represents the fractional excess of QSO–galaxy pairs found with a given separation over that which would be expected for a random distribution of the same total number of galaxies. In order to minimise random errors, for each separate QSO field (consisting of a 1 degree diameter region around the QSO) 100 times as many random 'galaxies' were generated as there were real galaxies in the field, and the counts then scaled appropriately.

When computing $w_{qg}(\theta)$, we excluded from the analysis a 10 arcsec radius region around each QSO. This ensured that we did not cross-correlate the position of the QSO with its own identification as a galaxy in the APM catalogue. We removed this large a region because Stocke et al. (1991) quote an error of $\pm 5$ arcsec on their optical position. That this radius was sufficient to prevent any self-correlation occuring was verified by inspection of each QSO field. This 'exclusion zone' also prevented any images caused by the break-up of the few bright ($B_J < 15$) EMSS QSOs from being included in the cross-correlation analysis. Conversely, since some of the low redshift objects are clearly extended, any loss of objects at small separations caused by images merging also does not affect our analysis.

The clustering signal from any single QSO is obviously very weak and has a large error associated with it, due to Poisson errors and the intrinsic clustering of faint galaxies. The advantage of a statistical approach to measuring the clustering, is that the signal from all the fields are combined, greatly reducing random counting errors. In the data presented here, we have therefore co-added the counts found in each bin for all the fields, as if all the fields of galaxies were overlaid on each other with their centres registered, before the pair counts were made. This amounts to the pair-weighted average of all the $j$ fields calculated individually, e.g.

$$w_{qg}(\theta) = \frac{\sum_{i=1}^{j} N_{QG_i}}{\sum_{i=1}^{j} N_{QR_i}} - 1.$$

For both low ($z < 0.3$) and high ($z > 0.3$) redshift samples, the fields around all Northern galactic hemisphere QSOs (hereinafter the Northern sample) and all Southern galactic hemisphere QSOs (the Southern sample) were processed separately and compared, as a consistency check for systematics between the two samples.

In the $z > 0.3$ sample we expect no visible correlation. At these redshifts the magnitude limit of the galaxy cata-logue ($B_J < 20.5$) corresponds to an absolute magnitude of $M_{B_J} = -19.9$ for an E/S0 galaxy, approximately 0.3 mag brighter than the 'knee' of the galaxy luminosity function, (K-correction and luminosity evolution taken from Metcalfe et al. 1991). We may then expect to detect the brightest cluster members for QSOs slightly above the $z = 0.3$ cut. To quantify the amplitude expected for the correlation between the $z > 0.3$ QSOs and the $B_J < 20.5$ galaxy sample we assumed that the QSOs are clustered in the same way as galaxies, and numerically integrated Limbers equation using smooth fits to the galaxy and QSO redshift distributions. The predicted cross-correlation amplitude for the $z > 0.3$ QSO sample is less than 5 per cent of the amplitude for the $z \leq 0.3$ QSO sample. This is clearly almost zero, but has been plotted in figure 2. Any appreciable correlation between $z > 0.3$ QSOs and $z < 0.3$ galaxies detected on the Schmidt plates, must therefore have been introduced during the data processing. Results obtained with this 'control' sample of $z > 0.3$ QSOs also provides a useful check on any selection effects which have governed the inclusion of QSOs into the EMSS catalogue. As a final check of the analysis method, the QSO samples were also cross-correlated against the positions of stars on the plate, which should produce the null result.

Errors were calculated by splitting each of the QSO samples (North and South, low and high redshift) into five subsets and calculating $w_{qg}(\theta)$ for each. The error quoted on each point is then the *rms* fluctuation in the five values derived.

It has become traditional to convert values of the amplitude of the correlation function $w_{qg}(\theta)$ at a fixed angular scale ($A_{qg}$) to the spatial covariance amplitude ($B_{qg}$), after Longair & Seldner (1979). This normalises such variables as the depth of data sampling and the variation in the relation between metric and angular size as a function of redshift, and makes it possible to compare amplitudes derived from separate surveys. Calculation of $B_{qg}$ is however, dependent on various other assumptions (Longair & Seldner 1979, Ellingson et al. 1991); most notably the galaxy luminosity function. The main aim of deriving $w_{qg}(\theta)$ is to compare its amplitude with that of the galaxy auto-correlation function, $w_{gg}(\theta)$. Given the similarity between the $n(z)$ relation for QSOs and galaxies in this survey, projection effects for $w_{qg}(\theta)$ will be similar to those for $w_{gg}(\theta)$. The relative strength of QSO and galaxy environments can therefore be determined by straightforwardly comparing the amplitudes of the two functions. In this case, the conversion from $A_{qg}$ to $B_{gq}$ becomes an unnecessary complication and source of uncertainty.

### 3.2 The integral constraint

As discussed above, the background level of galaxies was determined for each individual QSO field. We know however, from the auto-correlation of large catalogues of galaxies (e.g. Maddox et al. 1990a) that at scales of half a degree there is still appreciable clustering, leading to roughly a 5 per cent excess of galaxies over the true background density. By setting our background level too high, we weaken the apparent correlation. This is known as the 'integral constraint', which effectively forces the integral of $w_{qg}(\theta)$ over the angular scales studied to equal zero. This leads to a fractional error $\Delta B$ in the estimate of the background counts given by

$$\Delta B = \int 2\pi\theta w_{qg}(\theta)d\theta.$$

In order to calculate a value for $\Delta B$, it is necessary to substitute in an assumed form for $w_{qg}(\theta)$. We have used the angular galaxy auto-correlation function of Maddox et al. (1990a). Over the magnitude range $17 < B_J < 20.5$, the galaxy auto-correlation function is $w_{gg}(\theta) = 0.021 \times \theta^{-0.688}$, giving $\Delta B = 0.04046$. It must be stressed that this assumes a form for $w_{qg}(\theta)$ which has not been derived from our own data. However, we will show that this approach results in a QSO–galaxy correlation function which is indeed consistent with the galaxy auto-correlation function of Maddox et al. (1990a). Moreover, the level of the correction is small, only increasing the value of $w_{qg}(\theta)$ by $\sim 0.04$ over the angular scales studied. Despite being small, the integral constraint does mean that what looks to be zero correlation at separations larger than 500 arcsec will, with this correction become a small overall positive correlation.

## 4 RESULTS AND DISCUSSION

Table 1 contains the observed QSO–galaxy and QSO–random pair counts (before any correction for the integral constraint) found in each angular bin for the four data-sets used, along with the calculated $w_{qg}(\theta)$ and error. There is an excess of QSO–galaxy pairs over QSO–random pairs at small angular scales ($\theta < 4$ arcmin) for both the Northern and Southern $z < 0.3$ QSO samples. For completely independent pairs, the Poission significance of the excess is given by:

$$\sigma = \frac{(N_{QG} - N_{QR})}{\sqrt{N_{QR}}}.$$

The minimum separation between any two $z < 0.3$ QSOs is 9 arcmin and so all pair counts at $\theta < 4$ arcmin will comprise independent pairs. The Poisson significance for these first two angular bins is therefore $4.1\sigma$ for the Northern fields, and $2.6\sigma$ for the Southern fields. For the combined data set, we obtain $4.7\sigma$. On the basis of these results, we conclude that there is a statistical excess of galaxies around $z < 0.3$ QSOs.

For the QSOs at $z > 0.3$, $w_{qg}(\theta)$ is consistent with zero. For the combined Northern and Southern fields, the small excess detected in the first two bins has a formal significance of $0.13\sigma$. This is exactly as we would expect, and supports the genuine nature of the significant correlation observed in the low redshift range. Cross-correlations of both QSO redshift samples with stars also produced no significant signal. If $w_{qg}(\theta)$ is exactly zero, there is no integral constraint. If we assume that the very small correlation suggested above really is present, but not detected because of the noise, then an integral constraint of $\Delta B = 0.00194$ is applicable. We have not plotted this in any of our graphs.

Figure 2 plots $w_{qg}(\theta)$ for all 4 QSO samples, together with the $w_{gg}(\theta)$ for $B_J < 20.5$ galaxies from Maddox et al. (1990a). Although the error bars are large, it is clear that $w_{qg}(\theta)$ and $w_{gg}(\theta)$ show the same overall behaviour, but with an offset in amplitude. In figure 3 we present the pair-weighted average QSO–galaxy cross-correlation for all the $z \leq 0.3$ QSOs, after correction for the integral constraint. The agreement between the two correlation functions in this plot confirms our choice of $w_{gg}(\theta)$ to calculate the integral constraint and shows that the clustering of galaxies around QSOs is very similar to that of galaxies around other galax-ies. The chi-square best fit for the amplitude of $w_{qg}(\theta)$ is $A_{qg} = 0.0212$ (cf $A_{gg} = 0.021$), which provides an adequate fit to the data at the 99.99 per cent confidence level. Using a chi-square test, we place $2\sigma$ (i.e. 95 per cent confidence) limits on the amplitude of $w_{qg}(\theta)$ at one degree, assuming the $-0.688$ power law, of $0.013 \leq A_{qg} \leq 0.036$, i.e.

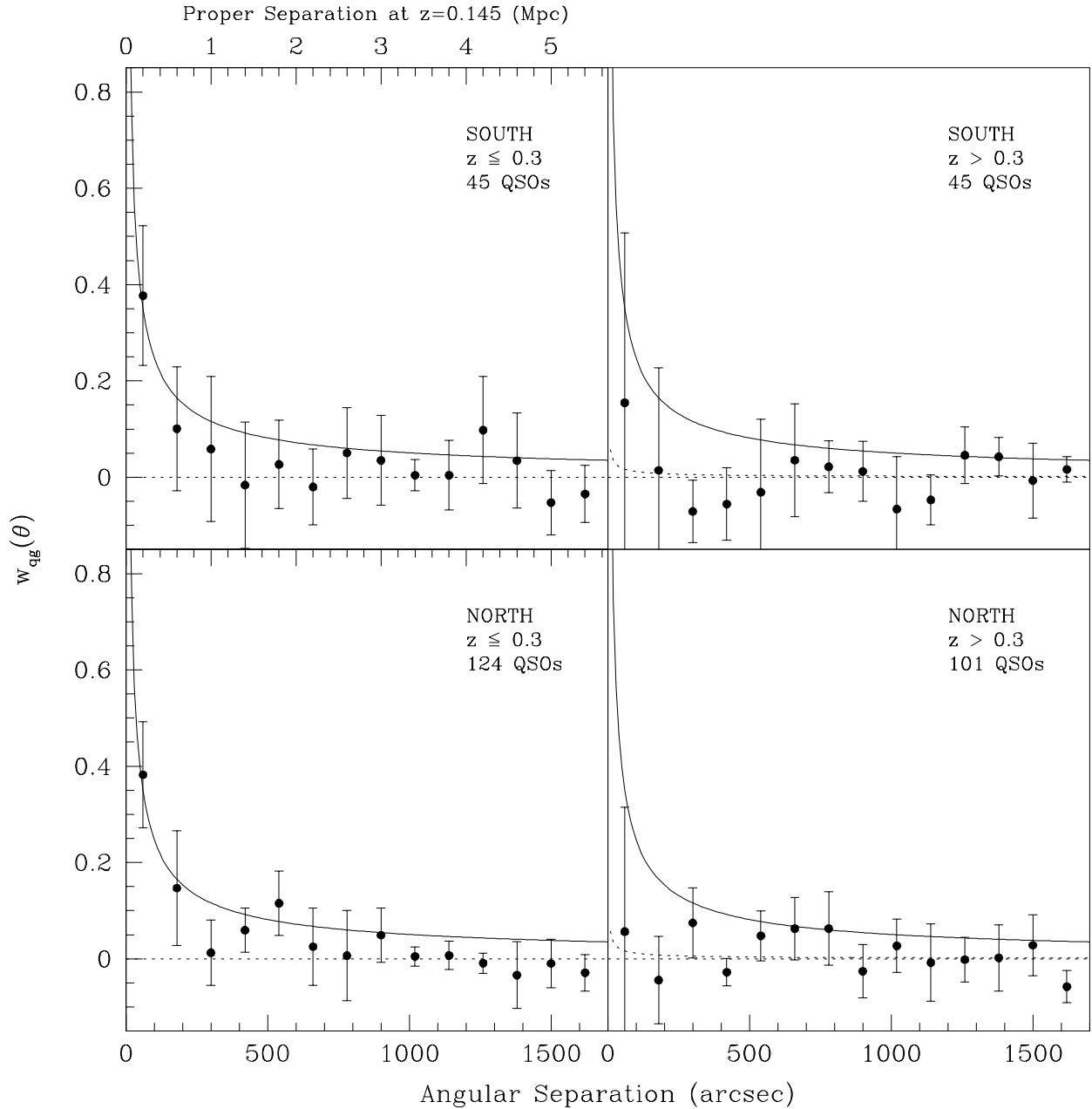

**Figure 2** The QSO–galaxy correlation function $w_{qg}(\theta)$ in the two ranges, $z \leq 0.3$ and $z > 0.3$, with the results for the Northern and Southern samples plotted separately. The solid curve is the galaxy–galaxy correlation function of Maddox et al. (1990a). The dotted curve shown for the $z > 0.3$ samples is the expected correlation, given that a few ($\sim 5$ per cent) of the galaxies are genuinely at these higher redshifts (see section 2), and assuming no evolution of the correlation amplitude. No account has been taken of the integral constraint in the estimation of $w_{qg}(\theta)$ (See section 3.2).

$$w_{qg}(\theta)/w_{gg}(\theta) = 1.0^{+0.7}_{-0.4}$$

to this depth and in the range $0 \leq z \leq 0.3$.

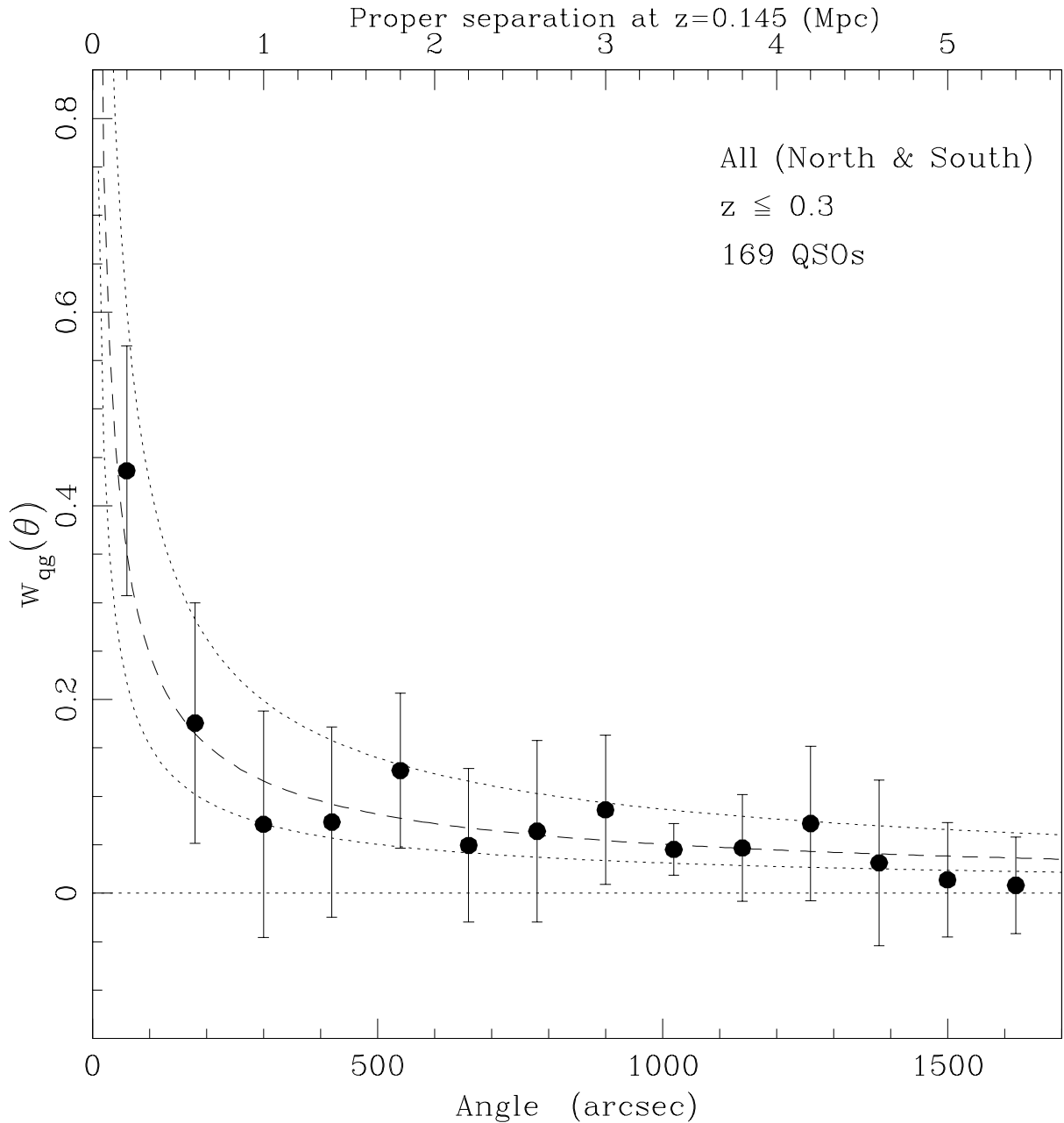

**Figure 3** The QSO–galaxy correlation function over the range $0 < z \leq 0.3$. The filled circles represent the pair-weighted average QSO-galaxy cross-correlation function for all the $z \leq 0.3$ QSOs, after correction for the integral constraint. The dashed curve is the galaxy–galaxy correlation function of Maddox et al. (1990a). The two dotted curves are the bounds of the 95 per cent confidence region for the amplitude of $w_{qg}(\theta)$ assuming the -0.688 power law. In calculating the background counts, the total number of galaxies present has been modified to take account of the integral constraint (See section 3.2).

As mentioned earlier, most results of this kind are presented in terms of the QSO–galaxy covariance amplitude $B_{qg}$. Values for the galaxy covariance amplitude $B_{gg}$ are in the range from 30 to 70 Mpc$^{1.77}$ (Koo & Szalay 1984). Ellingson et al. (1991) measure $\langle B_{qg}\rangle/\langle B_{gg}\rangle$ to be $1.1 \pm 0.6$ ($1\sigma$ errors) over the range $0.3 \leq z \leq 0.6$, assuming a galaxy covariance of 67.5 Mpc$^{1.77}$ (Seldner & Peebles 1978). Although Boyle & Couch (1993) do not detect any significant clustering around their optically-selected QSOs ($0.9 \leq z \leq 1.5$), their result $B_{qg} = 3 \pm 40$ Mpc$^{1.77}$ is also consistent with the measured galaxy covariances at the present epoch, and inconsistent with the very rich environment ($B_{qg} = 543 \pm 142 \text{Mpc}^{1.77}$) of radio-loud QSOs measured by Yee & Green (1987) at $z \sim 0.6$. Expressed in these terms, our result becomes $\langle B_{qg}\rangle/\langle B_{gg}\rangle = 1.0^{+0.7}_{-0.4}$ at the 95 per cent confidence limit. Combining this result with those of Ellingson et al. (1991) and Boyle & Couch (1993), we find that the radio-quiet QSOs studied show no appreciable evolution in the strength of their cluster environment over the redshift range $0 < z < 1.5$. This implies that radio-quiet QSOs are as unbiased a tracer of mass in the Universe as galaxies, but have the further advantages of sparse sampling and high luminosity.

The agreement between the QSO-galaxy and galaxy-galaxy correlation functions implies that both QSOs and galaxies are likely to have similar correlation lengths at the present epoch. This is consistent with the observed co-moving correlation length of QSOs at $z \sim 1.5$ (which is also the same as that of present day galaxies) only if the QSO correlation length is constant in co-moving separation (see Shanks & Boyle 1994). However, some care must be taken in interpreting this result, since the co-moving separations over which the correlation lengths are derived ($\leq 2$ Mpc for the QSO-galaxy correlation function at low redshifts and $\leq 10$ Mpc for the QSO-QSO correlation function at high redshifts), could correspond to different evolutionary regimes, i.e. non-linear and linear evolution respectively.

The magnitude distribution of this sample peaks at $M_V = -22.2$, a little fainter than what is sometimes considered to be the dividing line between QSOs and Seyfert 1 galaxies $M_V \sim -23$. Based on the evolution of the QSO optical luminosity function derived by Boyle et al. (1991), these objects are a magnitude fainter than the 'knee' of the two power law luminosity function at our mean redshift of $z = 0.15$. At $z = 2$, objects located at a similar position on the luminosity function would have $M_V \sim -26.8$ and would thus be considered bona fide QSOs. Indeed, objects 1 mag fainter than the 'knee' of the luminosity function form the bulk of $0.3 < z < 2.2$ QSOs identified in faint ($B_J < 21$) QSO surveys (Boyle et al. 1990). No significant difference is observed between the amplitude of $w_{qg}(\theta)$ determined separately for the higher luminosity ($M_V < -22.2$) and lower luminosity ($M_V > -22.2$) QSOs in the present sample.

## 5  CONCLUSIONS

We have obtained a significant ($4.7\sigma$) detection of ($B_J < 20.5$) galaxy clustering around a sample of 169 X-ray-selected QSOs. The amplitude of the QSO–galaxy cross-correlation function is $A_{qg} = 0.0212^{+0.015}_{-0.008}$, at one degree, where the errors represent the 95 per cent confidence limits. This is consistent with that of the galaxy–galaxy auto-correlation function ($A_{gg} = 0.021$) at the $B_J < 20.5$ magnitude limit. This result implies that these QSOs inhabit environments identical to those of normal galaxies. Coupled with existing observations of QSOs at higher ($z > 0.3$) redshifts, we also conclude that the environments of radio-quiet QSOs are consistent with no evolution over the redshift range $0 < z < 1.5$, in marked contrast to the rapid evolution in the richness of galaxy clusters associated with radio-loud QSOs.

## ACKNOWLEDGMENTS

We wish to thank Alastair C. Edge for helpful discussions on possible selection effects in the EMSS catalogue. RJS acknowledges the financial support of the Particle Physics and Astronomy Research Council.